\documentclass[a4paper,11pt]{article}
\pdfoutput=1 % if your are submitting a pdflatex (i.e. if you have
\usepackage{jinstpub} % for details on the use of the package, please
\usepackage{lineno}
\title{\boldmath Performance of a coarsely pixelated LAPPD photosensor for the SoLID gas Cherenkov detectors}

\author[a,1]{J. Xie,\note{Corresponding author.}}
\author[a]{C. Peng,}
\author[a]{S. Joosten,}
\author[a]{Z.-E. Meziani,}
\author[b]{A. Camsonne,}
\author[b]{M. Jones,}
\author[b]{S. Malace,}
\author[c]{E. Kaczanowicz,}
\author[c]{M. Rehfuss,}
\author[c]{N. Sparveris,}
\author[d]{M. Paolone,}
\author[e]{M. Foley,}
\author[e]{M. Minot,}
\author[e]{M. Popecki}
\author[f]{and Z.W. Zhao}

\affiliation[a]{Argonne National Laboratory, 9700 S Cass Ave., Lemont, IL 60439, USA }
\affiliation[b]{Thomas Jefferson National Accelerator Facility, 12000 Jefferson Ave., Newport News, VA 23606, USA }
\affiliation[c]{Department of Physics, Temple University, Philadelphia, PA 19122, USA }
\affiliation[d]{Department of Physics, New Mexico State University, Las Cruces, NM 88003, USA }
\affiliation[e]{Incom, Inc., 294 Southbridge Rd., Charlton, MA 01507, USA }
\affiliation[f]{Department of Physics, Duke University and
Triangle Universities Nuclear Laboratory, Durham, NC 27708, USA }
\emailAdd{jxie@anl.gov}

\abstract{The SoLID spectrometer's gas Cherenkov counters require photosensors that operate in a high luminosity and high background environment. The reference design features arrays of 9 or 16 tiled multi-anode photomultipliers (MaPMTs), distributed across 32 sectors, to serve the light-gas and heavy-gas Cherenkov counters, respectively. To assess the viability of a pixelated INCOM Large Area Picosecond Photodetector (LAPPD$^{\rm TM}$) as an alternative photosensor to replace MaPMT arrays in either detector, we evaluated its performance under realistic SoLID running conditions in Hall C at the Thomas Jefferson National Accelerator Facility (Jefferson Lab). 

The results of this test confirmed that the coarse-pixelated (2.5$\times$2.5 cm$^2$ pixel size) LAPPD is capable of handling the total projected signal and background rates of the three pillar SoLID experiments. The tested photosensor detected Cherenkov signals with the capability of separating single-electron events from pair production events while rejecting background. Although the design was not aimed at ring-imaging Cherenkov detectors, Cherenkov disk images were captured in two different gas radiators. Through a direct comparison with a GEANT4 simulation, we confirmed the experimental performance of the LAPPD.}


\begin{document}
\maketitle
\flushbottom

\section{Introduction}
\label{sec:intro}

The Solenoidal Large Intensity Device (SoLID) ~\cite{1,2,3} is a new spectrometer designed to exploit the full potential of the 12 GeV energy upgrade of the Continuous Electron Beam Accelerator Facility (CEBAF) at Jefferson Lab. The SoLID spectrometer will be operated at the full CEBAF luminosity of 10$^{39}$~cm$^{-2}$s$^{-1}$~\cite{3} in the Parity Violation Deep Inelastic Scattering (PVDIS) experiment and at a still high but reduced luminosity of 10$^{37}$~cm$^{-2}$s$^{-1}$ in the Semi-Inclusive Deep Inelastic Scattering (SIDIS) and Exclusive $J/\psi$ production near threshold experiments (the SIDIS-$J/\psi$ experiments).

The SIDIS-$J/\psi$ spectrometer design uses two Cherenkov counters for particle identification: a light gas Cherenkov (LGC) for e/$\pi$ separation and a heavy gas Cherenkov (HGC) for $\pi/K$ and $\pi/p$ separation. Hamamatsu flat panel multi-anode photomultipliers (MaPMTs) ~\cite{4} were selected as the baseline photosensors for these Cherenkov detectors. The MaPMTs have 64 pixels with a sensitivity down to signals from single photons. The signals from the pixels can be read out either individually (64 pixels) or in various summing schemes, e.g. summing 16 pixels into four quadrants. However, a major drawback of these sensors is their loss of efficiency in a magnetic field ~\cite{5,6}, requiring the use of magnetic shielding over the full photosensing area for their use in a magnetic field. An alternative sensor is the microchannel plate photomultiplier (MCP-PMT), a photosensor with a proven strong magnetic field resiliency, and high granularity. Its design consists of a stacked pair of MCPs in a chevron configuration for secondary electron amplification. Due to the significantly reduced electron amplification path, MCP-PMTs show an exceptional magnetic field tolerance~\cite{7}, a great advantage compared to the heavy and cumbersome mu-metal magnetic shielding needed for large array configurations of MAPMTs in the SoLID Cherenkov detectors. Unfortunately, the prohibitive cost of these commercial MCP-PMTs prevents their use in large photosensing arrays. 

Over the past decade a new type of MCP-PMT, the large area picosecond photodetector (LAPPD) was developed by the LAPPD collaboration ~\cite{8} and more recently commercialized by Incom, Inc ~\cite{9,10,11}. The LAPPDs apply resistive and secondary emissive layers on a large area of glass capillary through atomic layer deposition (ALD), achieving low-cost functional MCPs. As a result the final cost of a manufactured device is expected to be much lower per active photosensing area compared to traditional MCP-PMTs, meanwhile, the performance of the LAPPD is similar to that of the standard commercial MCP-PMTs, including timing resolution, position resolution, and dark count. The expected low cost and high performance make the LAPPD an ideal photosensor candidate for Cherenkov counters.  

To explore the potential of using LAPPDs as the photosensors for the SoLID Cherenkov counters and to evaluate its performance in a high-rate background environment, we performed a first opportunistic but limited test ~\cite{12} of a strip-line read-out Gen-I LAPPD at Jefferson Lab. Although the event separation between single-electron and pair-production events was poor due to a low quantum efficiency and non-pixelated readout, the results showed a promising performance of the GEN-I LAPPD in a harsh running environment of Hall C at Jefferson Lab. With the availability of Gen-II (capacitive coupling) pixelated LAPPD, a more comprehensive test was designed and performed to validate it as a viable photosensor for the SoLID gas Cherenkov counters as part of a pre-R\&D. In this paper, we describe the details of this comprehensive test, including the experimental setup, the Gen-II LAPPD geometry and the data taking procedure. Results are presented and discussed as we evaluate the potential of a coarse-pixelated LAPPD as a possible photosensor candidate for the SoLID gas Cherenkov counters. 

\section{Telescopic Cherenkov device design }
\label{sec:design}
\begin{figure}[htbp]
\centering % \begin{center}/\end{center} takes some additional vertical space
\includegraphics[width=1.0\textwidth,trim=0 0 0 0,clip]{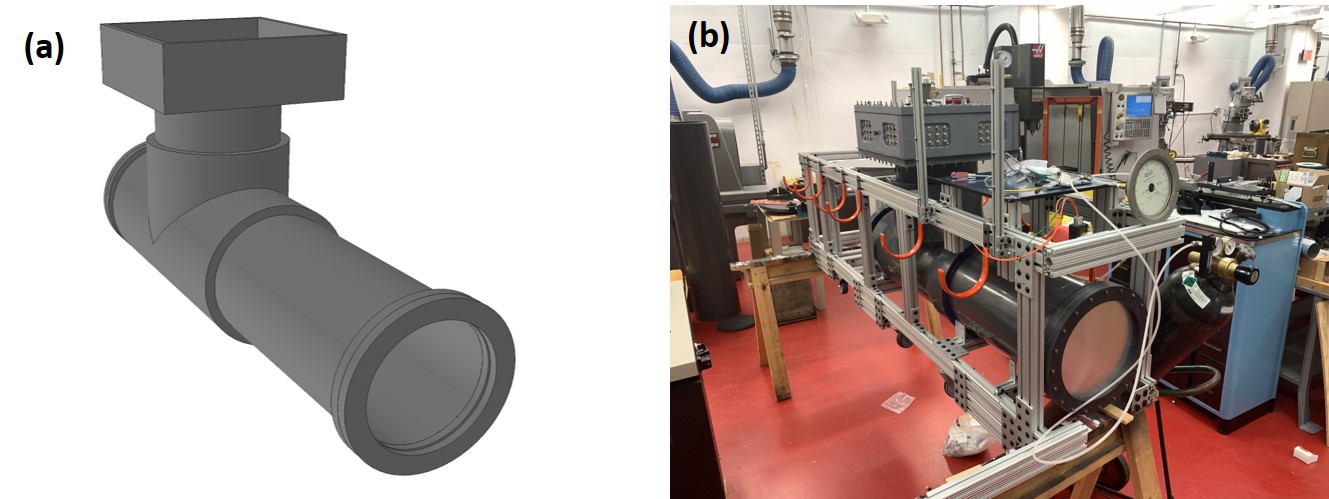}
\caption{\label{fig1} Schematic (a) and image (b) of the actual Cherenkov tank within its support structure as built at Temple University before being shipped to Jefferson Lab}
\end{figure}

A Telescopic Cherenkov Device (TCD) was designed and constructed, as shown in Figure~\ref{fig1}. Compared to Ref.~\cite{12} the present design allows the use of an array of 4$\times$4 MaPMTs or a 20$\times$20 cm$^2$ LAPPD, the size of the sensor array in the SoLID HGC. It consists of a primary Cherenkov tank, a LAPPD electronics housing box, an aluminum-coated reflective mirror, aluminum foil end caps, a gas supply system, a power supply system, and readout electronics. 

The primary Cherenkov tank was assembled from 14" diameter PVC pipes, and machine cut to the required dimensions. Inside the tank, felt flocking was attached to minimize secondary reflections. Additional apertures were added to allow valves and gas flow, sensors, and an LED test light source. A special housing box was designed to accommodate the coarse-pixelated LAPPD and its electronic cables. The housing box was designed to maintain structural and hermetic integrity while allowing detachable access plates in order to reach the enclosed electronics. To secure the removable LAPPD electronics housing box, large square rubber gaskets were applied. Aluminum windows and PVC window frames were pressure tested to ensure the system was airtight. Flat elliptical mirrors were constructed from carbon-fiber bases and Lexan reflective film. The primary mirror was mounted inside the tank and aligned to reflect light incident down the central ray of the tank and reflected by the mirror onto the center of the photon detector array. The LAPPD housing box was shielded by lead bricks during data taking to protect the photosensor from radiation damage. A 370 nm light-emitting diode (LED) was installed near the entrance window for calibration and monitoring purposes. The pressure regulation system included a gauge pressure manometer, two solenoid valves, a desktop PID controller with a laptop interface, and the required cabling. Operating slightly above atmospheric pressure helped simplify the mechanical design and minimize the entrance window thickness. Carbon dioxide (CO$_2$) and octafluorocyclobutane (C$_4$F$_8$) were used as radiators media to evaluate the LAPPD performance for both the light gas (CO$_2$) and heavy gas (C$_4$F$_8$) Cherenkov (see next section). In order to ensure high gas purity and before stabilizing it at atmospheric pressure, the telescopic Cherenkov device was flushed with 1 to 2 full detector equivalent volumes of the new gas each time the radiator was changed.

\section{Large area picosecond photodetector (LAPPD) }
\label{sec:lappd}

\begin{figure}[htbp]
\centering % \begin{center}/\end{center} takes some additional vertical space
\includegraphics[width=1.0\textwidth,trim=0 0 0 0,clip]{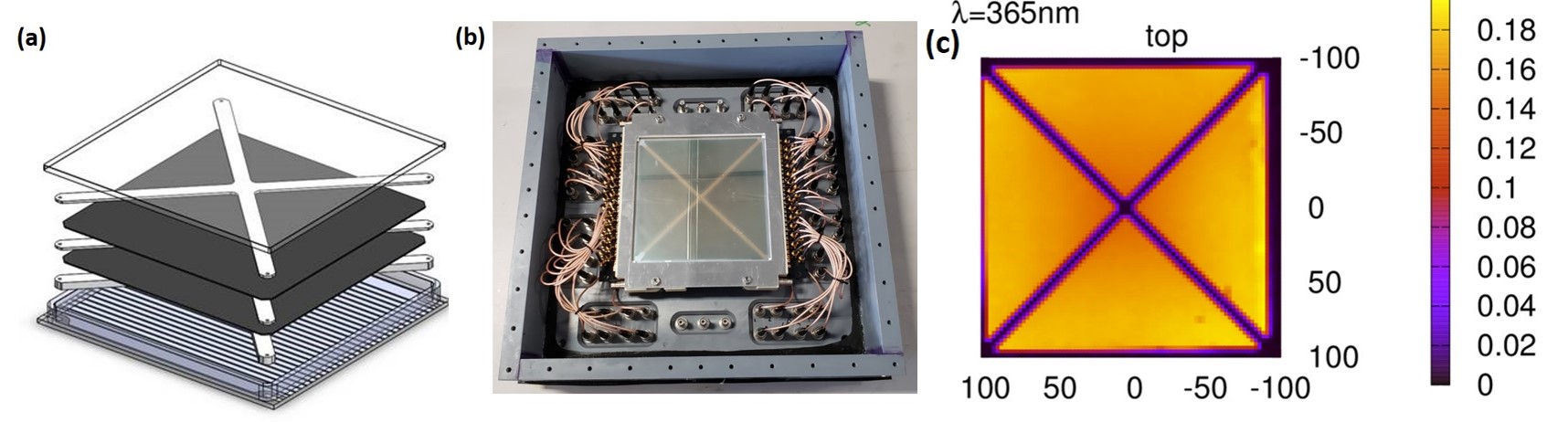}
\caption{\label{fig2} (a) Exploded-view drawing of a LAPPD; (b) Image of the 20$\times$20 cm$^2$ Gen II pixelated LAPPD installed in the LAPPD housing box; (c) Quantum efficiency map of the LAPPD at 365 nm, with an average of ~15\%.}
\end{figure}

The LAPPD photodetector is a novel microchannel plate photomultiplier based on the use of a newly developed microchannel plate, functionalized through atomic layer deposition on a glass capillary array structure (GCA-ALD-MCP) ~\cite{9}. Figure~\ref{fig2}(a) shows an exploded-view drawing of the LAPPD, consisting of a top window with an alkali photocathode on the inner surface to generate photoelectrons once incident photons enter the window, two microchannel plates with Chevron configuration to amplify the number of secondary electrons and a bottom plate to collect the resulting charge cluster for readout. 

The tested Gen-II LAPPD features a 64-pixelated readout with a pixel size of 25 mm $\times$ 25 mm through capacitive coupling. The entrance window of this LAPPD is regular B33 glass which absorbs photons below 270 nm. To enhance the Cherenkov photons detection efficiency at lower wavelengths, a thin layer of p-terphenyl wavelength shifter coating was applied~\cite{13} in order to convert ultraviolet photons into visible photons for detection. Fig.\ref{fig1} shows the exploded-view drawing of a LAPPD and an image of the tested coarse-pixelated LAPPD installed in the telescopic Cherenkov housing box, as well as its quantum efficiency map at 365 nm. The detailed parameters and performance on the bench of this particular LAPPD are listed in Table \ref{tab1}. 

\begin{table}[htbp]
\centering
\caption{\label{tab1} Geometry and performance parameters of the tested LAPPD}
\smallskip
\begin{tabular}{|l|l|}
\hline
  & Incom Gen-II LAPPD\\
\hline
PMT type &	MCP-PMT\\
Device size & 20~$\times$~20~cm$^2$\\
Active area  &	20~$\times$~20~cm$^2$\\
Readout type  &	Pixel (2.5cm $\times$ 2.5cm)\\
Entrance window material  &	B33 glass\\
Wavelength shifter coating  &	Yes \\
Response wavelength range  &	185 – 650 nm \\
Quantum efficiency (QE)  &	15\% (17\% maximum) \\
Operation high voltage  &  2350 V (MCP = 875 V)\\
Gain  &	9.5 × 10$^6$\\
\hline
\end{tabular}
\end{table}
	
\section{Experimental setup}
\label{sec:Exper}

The telescopic Cherenkov device was installed in Hall C at Jefferson Lab, along with the Super High Momentum Spectrometer (SHMS). The experiment was performed parasitically with the ongoing Hall C fixed-target experiments, the setup location was chosen for minimal interference with the running experiment as well as to have a moderate rate background environment. The Cherenkov was positioned to face the liquid hydrogen target to allow the scattered electrons from the target to freely pass through the Cherenkov entrance window after crossing the target side walls and the target scattering chamber aluminum walls.   
	
\begin{figure}[htbp]
\centering % \begin{center}/\end{center} takes some additional vertical space
\includegraphics[width=0.5\textwidth,trim=0 0 0 0,clip]{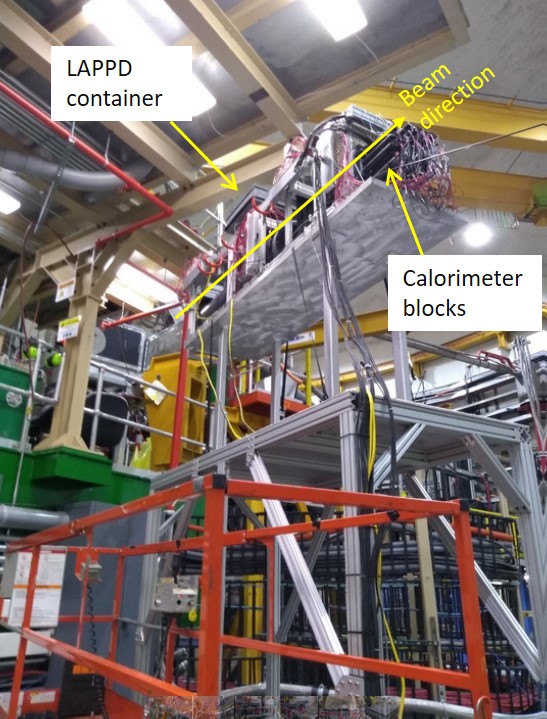}
\caption{\label{fig3} Telescopic Cherenkov device installed on a platform in Hall C at Jefferson Lab. The setup was lifted to the level of the beamline, with the entrance window facing the target scattering chamber. Scintillator paddles and calorimeter blocks were used at the back for forming a trigger. The LAPPD was protected from radiation damage by surrounding its housing with lead blocks.}
\end{figure}

The experiment used a 10.4 GeV electron beam, impinging on a polarized gaseous $^3$He target. While the scattered electrons or produced particles passed through the tank, Cherenkov light is produced along the particle path if its velocity fulfills the Cherenkov light emission condition. The Cherenkov light is then reflected at 90$^{\circ}$ by the mirror at the ${\rm T}$ section, and collected by the LAPPD. The coarse-pixelated LAPPD was accommodated in the housing box, and surrounded by lead blocks to prevent radiation damage. Scintillator paddles and Shashlik electromagnetic calorimeter blocks~\cite{14} were placed in front of the entrance window and behind the exit window respectively to generate coincident external triggers. All triggers and signals from the scintillator paddles, the calorimeter, and LAPPD were fanned out and sent to F250 flash Analog-to-Digital Converters (FADC250)~\cite{15} for data acquisition.  

The current SoLID detector design includes two Cherenkov counters: the LGC and the HGC. The LGC uses CO$_2$ at 1 atm (index of refraction n = 1.00045) as a gas radiator, and will be mainly used to discriminate electrons (17 MeV Cherenkov light threshold) from pions (4.75 GeV Cherenkov light threshold). The HGC, critical for the SoLID SIDIS program at forward angles from 8$^{\circ}$ to 15$^{\circ}$, uses C$_4$F$_8$ at 1.7 atm (index of refraction n = 1.0014) as a gas radiator and provides a clear separation between charged pions and kaons Cherenkov signals in a momentum range between 2.5 GeV/c and 7.5 GeV/c . To evaluate the performance of pixelated LAPPD in both the light gas and heavy gas Cherenkov, we experimented with CO$_2$ and C$_4$F$_8$ as the gas radiators, respectively. We conducted these tests in addition to rate dependence studies with the baseline Ma-PMT arrays.

The Cherenkov counter was installed in Hall C at Jefferson Lab, as shown in Figure~\ref{fig3}. It was mounted at beamline level pointing towards the target cell and positioned 5.4~m away. Its entrance window was facing the target at an angle of 75$^{\circ}$ with respect to the beam direction. The distance and angle were constrained by space and safety considerations due to the ongoing experiment in the hall, resulting in a moderate background-rate environment. Lead blocks shielded the photosensors from direct radiation damage.

We exposed the pixelated LAPPD to rates of up to 21 kHz per pixel, consistent with the projected rates for the SoLID SIDIS-$J/\psi$ conditions. During the same test period, we evaluated the MaPMT arrays at smaller angles under rate conditions with 0.25 photoelectron rates exceeding 3.4 MHz per quadrant, surpassing the projected single photon background rate of 7.1 MHz per MaPMT for the SoLID-PVDIS configuration. Previous studies on strip-line LAPPDs~\cite{12} have demonstrated their robustness in high-rate environments, suggesting that LAPPDs, like MaPMTs, are suitable for the harsh PVDIS conditions. Consequently, LAPPDs present a viable alternative to MaPMTs for the SoLID gas Cherenkov detectors.

\section{Simulation }
\label{sec:Simulation }

\begin{figure}[htbp]
\centering % \begin{center}/\end{center} takes some additional vertical space
\includegraphics[width=0.65\textwidth,trim=0 0 0 0,clip]{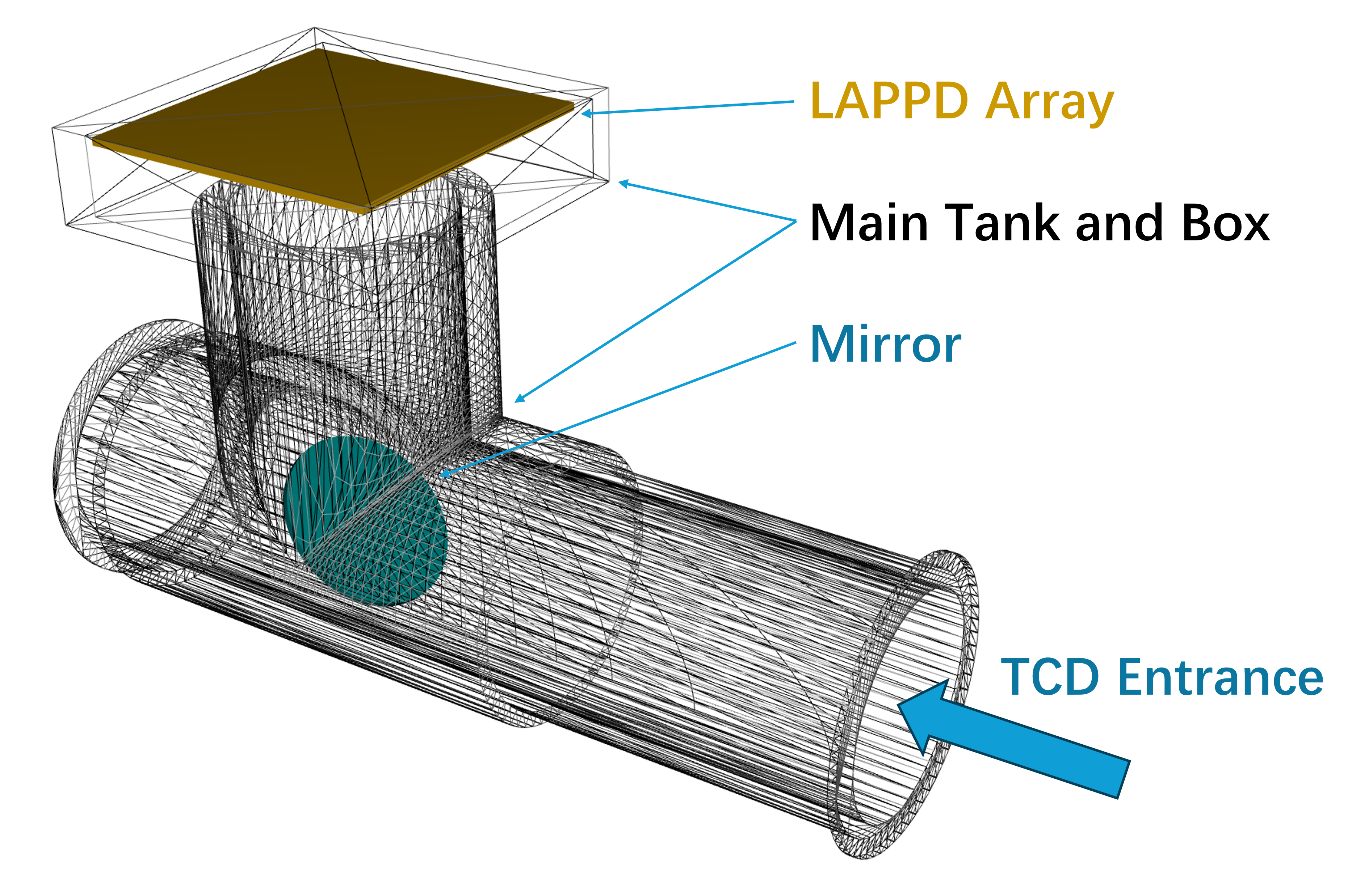}
\caption{\label{fig4} 
The Telescopic Cherenkov Detector setup in the GEANT4 simulation. The tank and box are drawn in wireframes to reveal the mirror and the LAPPD array.}
\end{figure}

In order to understand the experimental test results and later optimize the SoLID  LGC  design, we conducted a thorough simulation of the entire Cherenkov process.
A Geant4 simulation ~\cite{16} of the TCD, including the Cherenkov counter, the EM calorimeter (EC), and scintillator (SC) detectors, was developed similar to the SoLID simulation. The TCD setup in our simulation is shown in Figure~\ref{fig4}. 
The measured LAPPD quantum efficiency was used and a conservative Cherenkov optical physics cut-off was chosen to be 200 nm. 

We studied the resulting number of photo-electrons in the Cherenkov detector from two main sources, namely high energy scattered electrons and $\pi^0$'s production with subsequent decay photons conversion into electron-positron pairs from the polarized $^3$He gas target. Simulation results are presented together with the experimental test results for comparison in the following section.

% The TCD setup in the Geant4 simulation.

\section{Results and Discussion}
\label{sec:Results}

\begin{figure}[htbp]
\centering % \begin{center}/\end{center} takes some additional vertical space
\includegraphics[width=0.75\textwidth,trim=0 0 0 0,clip]{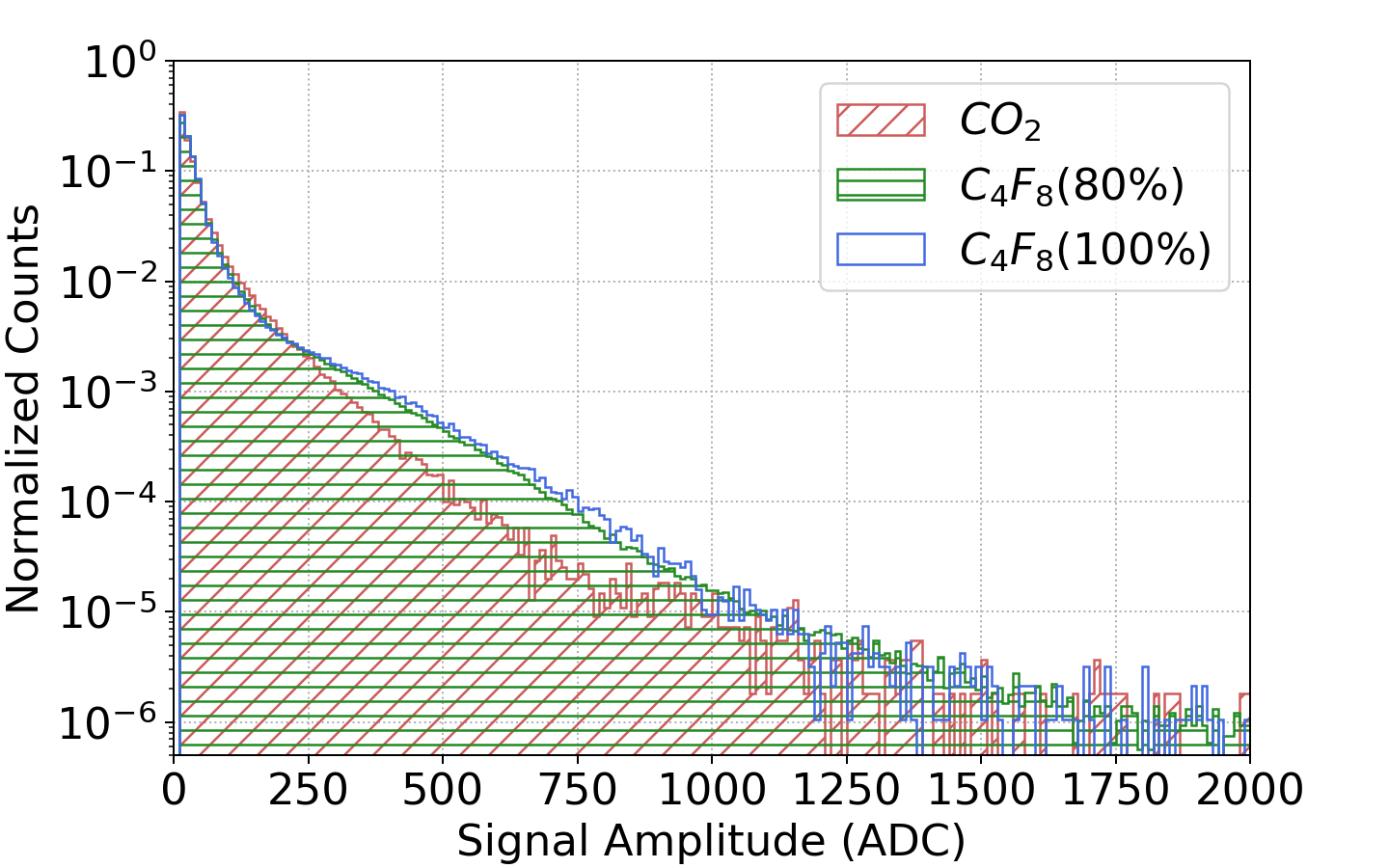}
\caption{\label{fig:signal_sum} Signal sum of all 64 LAPPD pixels for different gas radiators from the beam test. }
\end{figure}

The LAPPD utilizes a readout scheme featuring an array of 8$\times$8 pixels, each reading a 25 mm $\times$ 25 mm area. Throughout the beam tests, various radiator gas mixtures were employed, including CO$_2$, an 80\% C$_4$F$_8$ / 20\% CO$_2$ blend, and pure C$_4$F$_8$. Signals from all 64 pixels were recorded, and the analysis was conducted offline. Figure~\ref{fig:signal_sum} illustrates the normalized distributions of the aggregated signal across all 64 pixels. A coincidence timing cut of $\Delta t = \pm 10$ ns has been applied to the relative timing between pixels and the triggering calorimeter block. All three distributions exhibit a consistent peak below 100 ADC channels, indicating a stable background of random coincidences arising from the high-rate environment. Notably, signals in the heavier gas, C$_4$F$_8$, show a prominent signal amplitude bump between ADC channel 500 and 750, most likely a result of the Cherenkov radiation in C$_4$F$_8$.

The single photoelectron (SPE) responses were analyzed using data collected outside the timing coincidence window. This analysis revealed that the SPE peak consistently ranged between ADC channels 16 and 20 across all pixels. Given that the signal amplitude for the majority of our data does not exceed 2 SPE, we assumed a linear response. Consequently, all signals were normalized to the SPE amplitude to extract the number of photoelectrons (NPE).

As depicted in Figure~\ref{fig:signal_sum}, the data are overwhelmed by the random coincidences background at low ADC channels, an expected phenomenon caused by the high-rate environment experienced during the test. Thus, distinguishing Cherenkov signals from this background presents a significant challenge, particularly for the light gas radiator (CO$_2$) because of its relatively low light yield. To effectively select the Cherenkov signals, it is imperative to leverage the granularity afforded by the pixelized readout. Typically, photons from a Cherenkov event are distributed in an annular pattern on the photosensor plane, illuminating multiple pixels simultaneously. The exact number of activated pixels varies depending on the photon annulus's location and size, which are affected by the incident particle's angle and momentum, the refractive index of the radiator, and the photon's path length to the photosensors.

\begin{figure}[htbp]
\centering % \begin{center}/\end{center} takes some additional vertical space
\includegraphics[width=0.85\textwidth,trim=0 0 0 0,clip]{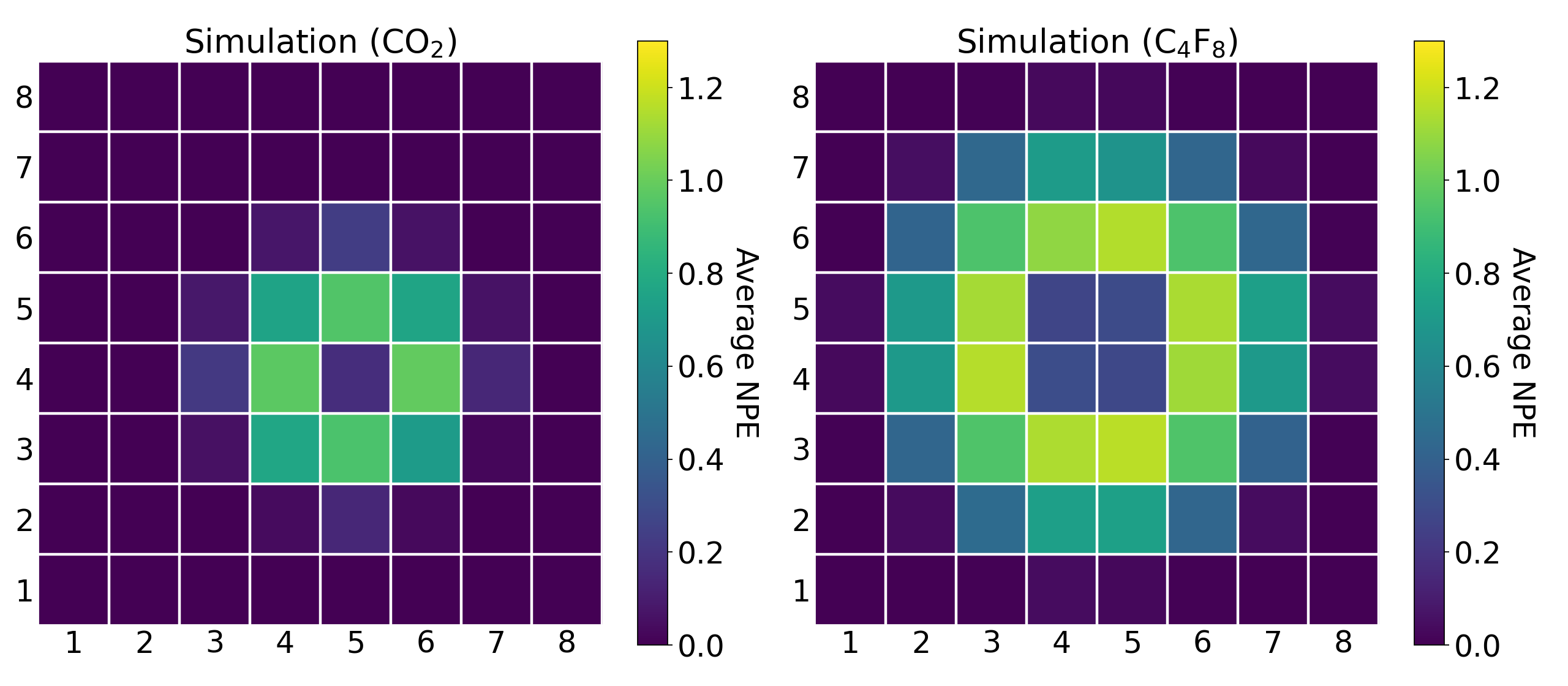}
\caption{\label{fig:annulus_shape_sim}
Simulation of the average NPE per event for the LAPPD beam test with different gas radiators. In C$_4$F$_8$ (right), the incoming electrons enter the TCD along its centerline. In CO$_2$ (left), the electron trajectories are shifted by a small amount so the Cherenkov photons are centered at pixel $P_{55}$.}
\end{figure}

The simulation results for Cherenkov photon distributions are shown in Figure~\ref{fig:annulus_shape_sim}. For CO$_2$ gas (with a refractive index $n_{CO_2}$ = 1.00045), the photons annulus covers 12 pixels when they are centered at one pixel, activating roughly 8 pixels given the LAPPD's quantum efficiency. In the case of C$_4$F$_8$ gas radiator ($n_{C_4F_8}$ = 1.0014), the Cherenkov light cone contains a higher photons count and exhibits a larger opening angle, resulting in a coverage of 32 pixels and approximately 15 activated pixels if the trajectories of the incident electrons coincide with the TCD's centerline. Here a pixel is considered "activated" if its signal exceeds 0.5 NPE, which is an arbitrary choice to reject dark current signals. Our results are stable by varying this threshold from 0.4 to 0.6 NPE.

\begin{figure}[htbp]
\centering % \begin{center}/\end{center} takes some additional vertical space
\includegraphics[width=0.85\textwidth,trim=0 0 0 0,clip]{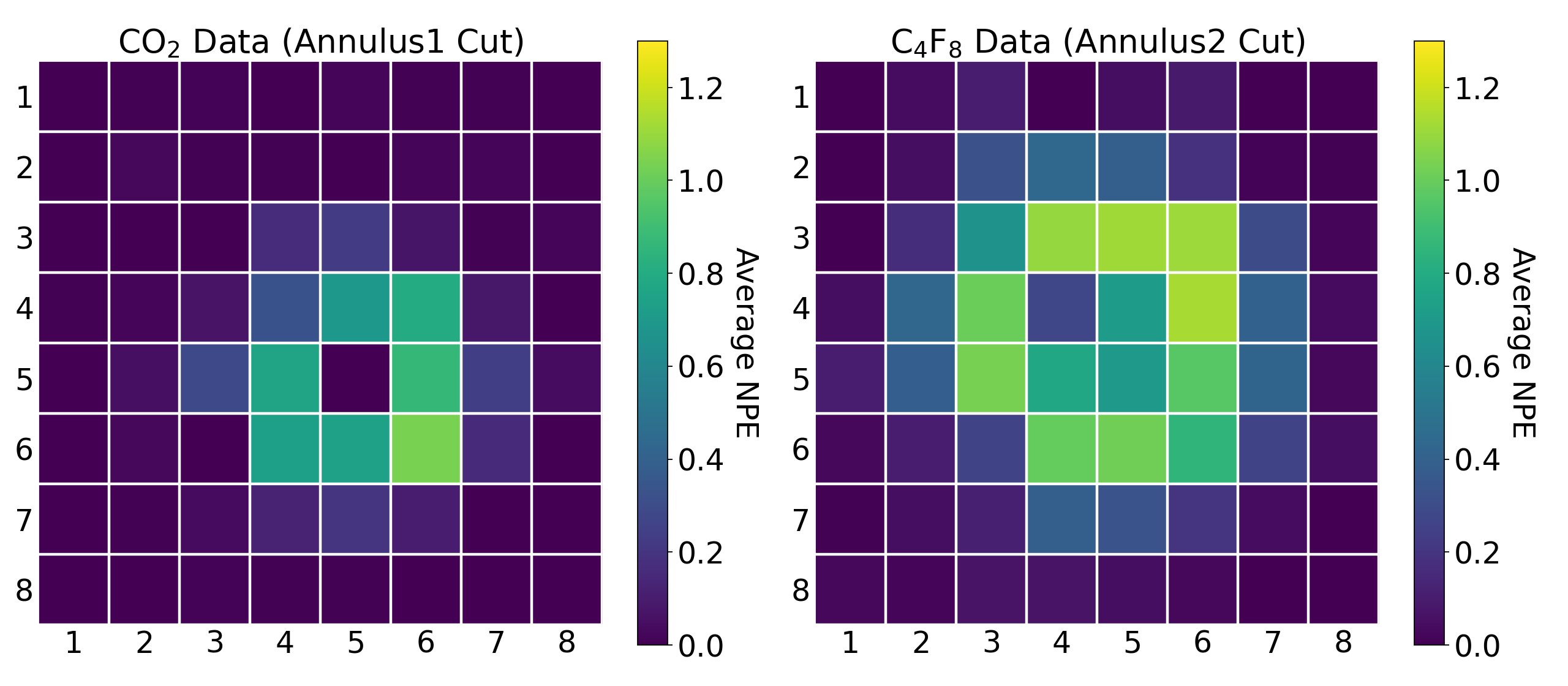}
\caption{\label{fig:annulus_shape_data}
Average NPE distributions with the ``Annulus1'' cut on the CO$_2$ data (left) and the ``Annulus2'' cut on the C$_4$F$_8$ data (right). Both cuts are centered at $P_{55}$.}
\end{figure}

The shape of the Cherenkov photon annulus is also studied with beam test data. To select clean samples of Cherenkov events, square annulus cuts were applied to the data. The annulus is one pixel wide and its center is defined as the bottom right corner of the inner square. The cut can be applied on the center pixel $P_{ij}$ (at \textit{i}th row and \textit{j}th column of the array) and its surrounding pixels with three parameters: the size of the inner square ($\rm{N_{inner}}$), the number of inactivated pixels inside the inner square ($\rm{N_{inact}^{inner}}$), and the number of activated pixels within the annulus ($\rm{N_{act}^{annulus}}$). Table \ref{tab:ring_cut_table} lists two sets of parameters chosen for different gas radiators. The photon distribution with the annulus cuts on $P_{55}$ for both CO$_2$ and C$_4$F$_8$ data are shown in Figure~\ref{fig:annulus_shape_data}. The observed size and shape of the photon annulus in the data agree well with the predictions from our simulations.

\begin{table}[h!]
\centering
\caption{\label{tab:ring_cut_table} Annulus cut parameters for selecting Cherenkov photons. The cut center is defined as the bottom right corner of the inner square.}
\smallskip
 \begin{tabular}{| c c | c c c c |} 
 \hline
 Cut & Gas Radiator & Center $P_{ij}$ & $\rm{N^{inner}}$ & $\rm{N_{inact}^{inner}}$ & $\rm{N_{act}^{annulus}}$ \\ [0.5ex] 
 \hline
 Annulus1 & CO$_2$ & $2 \leq i, j \leq 7$ & $1\times1$ & $1$ (out of 1) & $\geq 5$ (out of 8) \\ 
 Annulus2 & C$_4$F$_8$ & $3 \leq i, j \leq 7$ & $2\times2$ & $\geq 2$ (out of 4) & $\geq 8$ (out of 12) \\
 \hline
 \end{tabular}
\end{table}

\begin{figure}[htbp]
\centering % \begin{center}/\end{center} takes some additional vertical space
\includegraphics[width=0.99\textwidth,trim=0 0 0 0,clip]{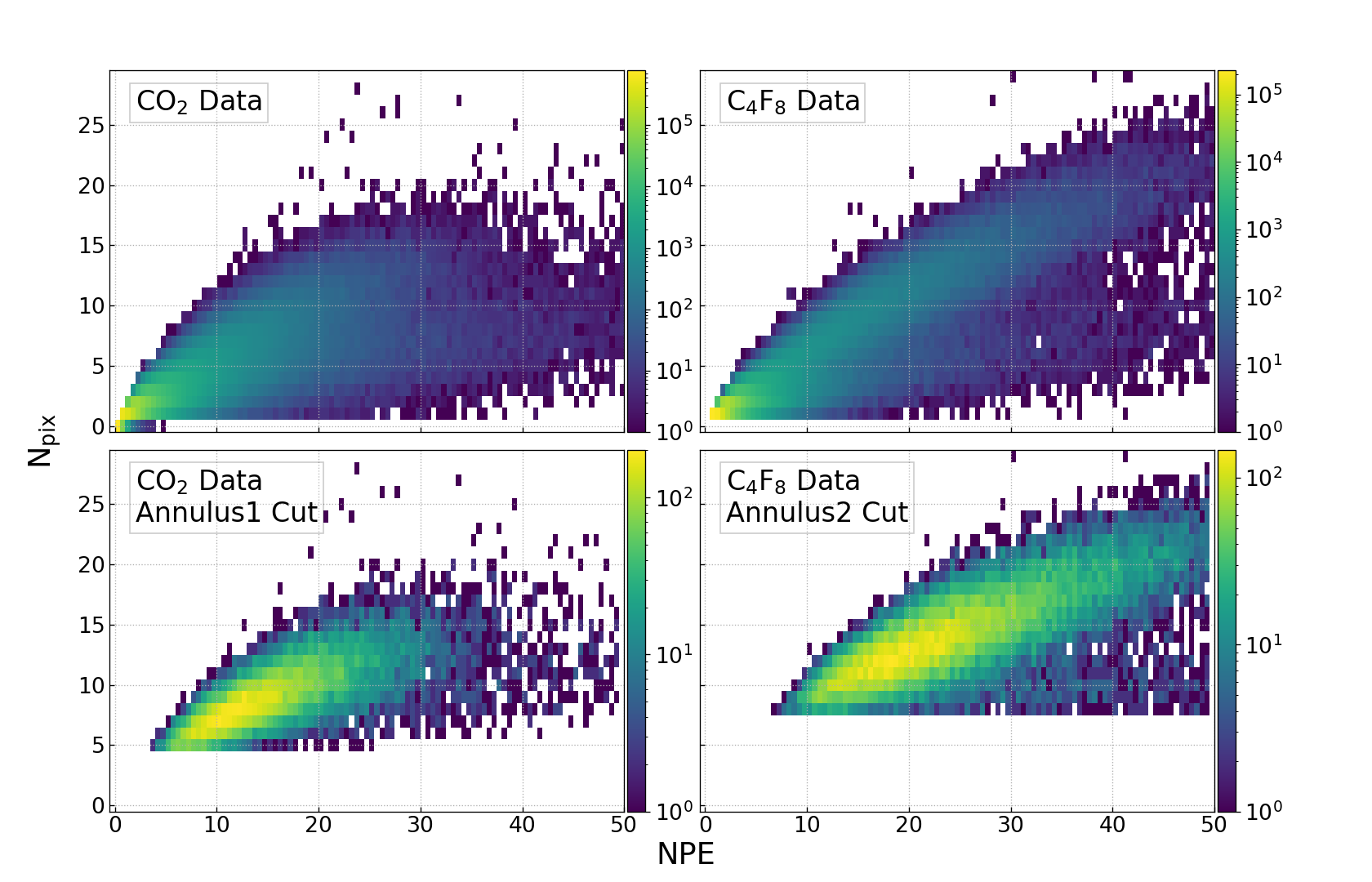}
\caption{\label{fig:sum2d_cuts} 2D distributions of signal sums for beam test data with CO$_2$ (left) and C$_4$F$_8$ (right) gas radiators.
Top: Events after a trigger timing cut ($|\Delta t| <= 10$ ns). Bottom: Events after the trigger timing cut and an annulus shape cut defined in Table \ref{tab:ring_cut_table}.}
\end{figure}

The annulus cuts are then applied across all the possible centers (see Table~\ref{tab:ring_cut_table}) on the photosensor to improve the statistics of the Cherenkov event samples. Figure~\ref{fig:sum2d_cuts} shows the 2D distribution of the selected events over the NPE and the number of activated pixels (N$_{pix}$). For both CO$_2$ and C$_4$F$_8$ data, a clear peak is revealed after the cut, showing that a typical Cherenkov event would activate up to 11 pixels in CO$_2$ and 17 pixels in C$_4$F$_8$ in the TCD.

\begin{figure}[htbp]
\centering % \begin{center}/\end{center} takes some additional vertical space
\includegraphics[width=0.85\textwidth,trim=0 0 0 0,clip]{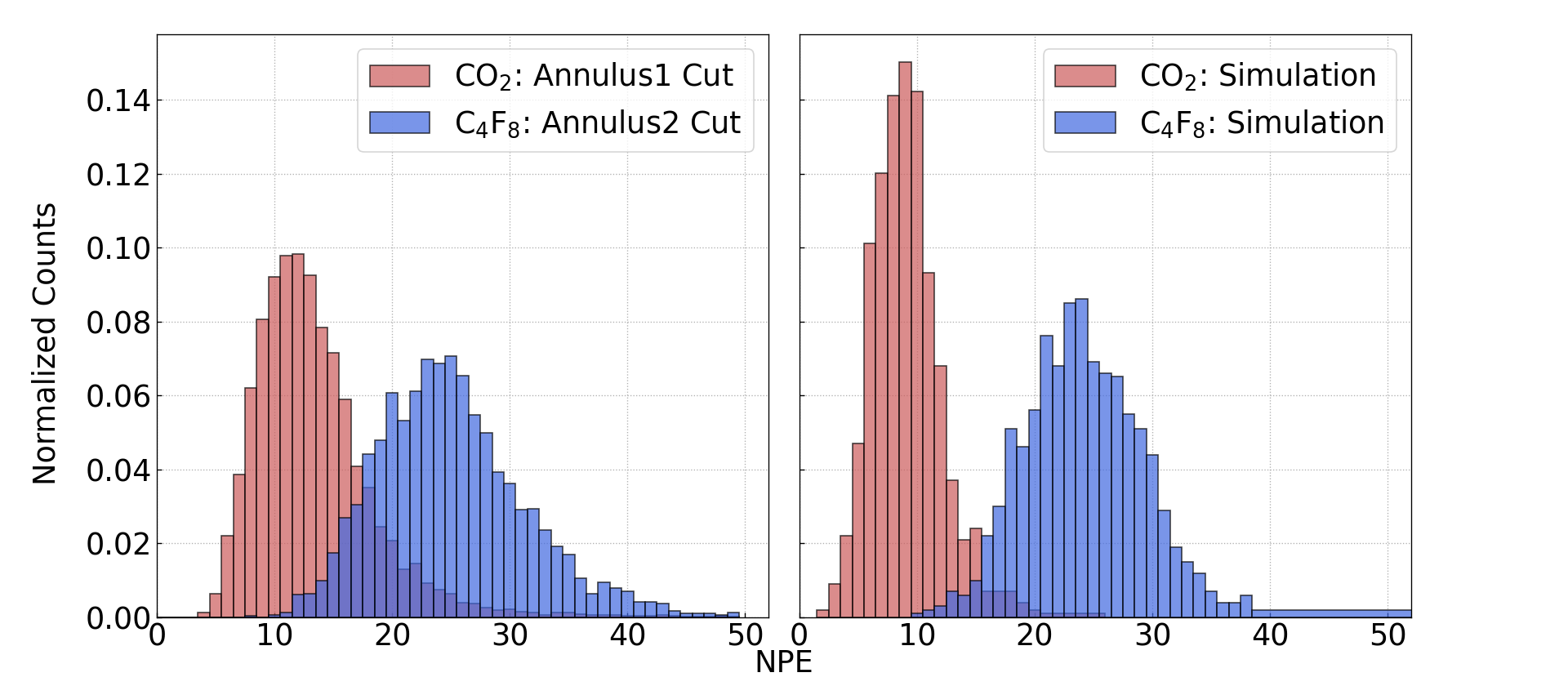}
\caption{\label{fig:npe_comparison}
Comparison between Cherenkov signals from CO$_2$ and C$_4$F$_8$. Left: Data taken with the LAPPD at a large angle. Annulus cuts have been applied to select Cherenkov events. Right: Simulation for the LAPPD response at a large angle with scattered electrons entering along the centerline of the detector.}
\end{figure}

The NPE distributions from the data and the simulation are shown in Figure~\ref{fig:npe_comparison}. The same annulus cuts are used for the beam test data. In addition, we also require the number of activated pixels outside the annulus shape to be less than that on the annulus, which significantly suppresses the large NPE tail on the distribution plot. This distribution characterizes the LAPPD responses to the Cherenkov photons in TCD, giving $\rm{NPE} = 12$ for CO$_2$ and $\rm{NPE} = 24$ for C$_4$F$_8$. The simulation also describes the difference between the two gas radiators, yielding an NPE of 9 for CO$_2$ and 23 for C$_4$F$_8$. The beam test data and simulation are consistent with each other on the NPE distribution of the Cherenkov signals.

\section{Summary}

We report on the performance study of a coarse-pixelated LAPPD using a telescopic Cherenkov device in an open high-rate environment at Jefferson Lab. A pixelated Gen-II LAPPD was tested with two different radiators, a light gas CO$_2$ and a heavy gas C$_4$F$_8$, which are planned to be used in the SoLID experiment. We demonstrate that the coarse-pixelated Gen-II LAPPD device can operate in a high-rate environment, consistent with the expected rates during the SoLID $J/\psi$ and SIDIS experiments. Furthermore, the pixelated information is critical to selecting a clean sample of the Cherenkov events from the random coincidences background. The LAPPD's response to Cherenkov photons is studied with two gas radiators, namely CO$_2$ and C$_4$F$_8$. As expected, more photoelectrons were generated from the Cherenkov process in C$_4$F$_8$ compared to those in CO$_2$. GEANT4 simulations were conducted to crosscheck the results, showing a good agreement with the data. Our experiment confirms that the pixelated Gen-II LAPPD is capable of handling the high-rate environment expected for future SoLID experiments. 

\section*{Acknowledgement}

Argonne National Laboratory's work was supported by the US Department of Energy, Office of Science, Office of Nuclear Physics, under contract DE-AC02-06CH11357. The work at Temple University is supported by the US Department of Energy award DE-SC0016577. Incom's work is supported by the U. S. Department of Energy, Office of Science, Office of Basic Energy Sciences, Offices of High Energy Physics and Nuclear Physics under DOE contracts: DE-SC0015267, DE-SC0017929, DE-SC0018778, and DE-SC0019821. The work at Duke University is supported by the US Department of Energy under contract DE-FG02-03ER41231. This material is based upon work supported by the U.S. Department of Energy, Office of Science, Office of Nuclear Physics under contract DE-AC05-06OR23177.

\end{document}